
\input harvmac.tex

\def\inbar{\,\vrule height1.5ex width.4pt depth0pt}
\def\IB{\relax{\rm I\kern-.18em B}}
\def\IC{\relax\hbox{$\inbar\kern-.3em{\rm C}$}}
\def\IP{\relax{\rm I\kern-.18em P}}
\def\IR{\relax{\rm I\kern-.18em R}}
\font\cmss=cmss10 \font\cmsss=cmss10 at 7pt
\def\IZ{\relax\ifmmode\mathchoice
{\hbox{\cmss Z\kern-.4em Z}}{\hbox{\cmss Z\kern-.4em Z}}
{\lower.9pt\hbox{\cmsss Z\kern-.4em Z}}
{\lower1.2pt\hbox{\cmsss Z\kern-.4em Z}}\else{\cmss Z\kern-.4em Z}\fi}

\def\trace{{\rm tr~}}
\def\half{{1\over2}} 
\def\cint{\oint}

{\nopagenumbers\abstractfont\hsize=\hstitle%
\rightline{\vbox{\baselineskip12pt\hbox{RU-93-13}\hbox{NSF-ITP-93-39}}}%
\vskip .3in\centerline{\titlefont%
\vbox{\centerline{Conformal Field Theory Techniques}}}%
\vskip .2in\centerline{\titlefont%
\vbox{\centerline{for Large $N$ Group Theory}}}%

\abstractfont\vskip .3in\pageno=0}

\centerline{Michael R. Douglas$^1$}
\centerline{Dept. of Physics and Astronomy}
\centerline{Rutgers University}
\centerline{\it and}
\centerline{Institute for Theoretical Physics}
\centerline{University of California at Santa Barbara}
\bigskip
\bigskip
\bigskip
\noindent
We show how to use quantum mechanics on the group manifold $U(N)$
as a tool for problems in $U(N)$ representation theory.
The quantum mechanics reduces to free fermions on the circle, which in the
large $N$ limit become relativistic.  The theory can be bosonized giving
the
Das-Jevicki-Sakita collective field theory.
The formalism is particularly suited to problems involving tensor product
multiplicity (Littlewood-Richardson) coefficients.
As examples, we discuss the partition function of two-dimensional
Yang-Mills
theory on the sphere, and the zero magnetic field limit of $D$-dimensional
Eguchi-Kawai Yang-Mills theory.
We give the leading $O(N^0)$ solution of the latter theory, using a method
which allows computing corrections.

\footnote{}{$^1$ (mrd@physics.rutgers.edu, or \ @rutgers.bitnet)}

\Date{March 29, 1993}

In the following we describe some techniques for studying group
representations
of $U(N)$ and $SU(N)$ based on those that have been used recently to study
the
matrix model of $D=1$ strings.
\ref\gm{For a comprehensive review, see I. Klebanov's April 1991 Trieste
lectures, or the upcoming work of P. Ginsparg and G. Moore.}~~
Some aspects of this are already described in
\ref\jev{A. Jevicki, Nucl. Phys. B376 (1992) 75}
and references there.
Others are in the mathematics literature
\ref\ps{A. Pressley and G. Segal, Loop Groups, Clarendon, Oxford, 1986}
and similar ideas have also been applied in the theory of the quantum Hall
effect.  \ref\stone{M. Stone, Phys. Rev. B42 (1990) 8399}
Here we will concentrate on applying them to large-$N$ Yang-Mills theory.
A more complete account will appear in upcoming work.

\nref\mig{A.  Migdal, Zh.  Eksp.  Teor.  Fiz. {69} (1975) 810
 (Sov.  Phys.  JETP.  {42} 413)}%
\nref\rus{B. Rusakov, Mod. Phys. Lett. {A5} (1990) 693.}%
The immediate motivation for the present work was the study of
two-dimensional
Yang-Mills theory, which was solved exactly in \refs{\mig,\rus}.
The solution is presented as a sum over group representations; for a genus
$G$
target space of area $A$, with 2d Yang-Mills coupling $g$,
\eqn\qcd{Z_G(g^2 A) = \sum_R (\dim R)^{2-2G} e^{-g^2 A C_2(R)}.}
Performing such sums in the large $N$ limit with $g^2 N$ fixed is not
trivial.
In recent work D. Gross and W. Taylor
\ref\gt{D. Gross and W. Taylor, preprints PUPT-1376 hep-th/9301068 and
PUPT-1382 hep-th/9303046.}
have developed a string representation
which represents the series as a sum of terms
$\sum_{m,n} w_{mn} (g^2 A)^m e^{-n g^2 A}$, whose weights $w_{mn}$ are
determined by a sum over maps with a specified winding number $n$ and
having
$m$ distinguished points.  For $G\ge 1$ this sum has a relatively simple
form,
while for $G=0$ the sum is infinite and has so far resisted exact
evaluation.
We reformulate it here in a way that we believe will be more tractable.

Although certain objects natural in QCD${}_2$, such as the ``three-space''
vertex, seem significantly more complicated in this formalism, there are
other
operations which are much simpler in this description than in other group
theory techniques.
A prime example is multiplication of characters, which can be used to
compute
tensor product characters and multiplicity (Littlewood-Richardson)
coefficients
for irreducibles in tensor products.
Using this operation, we formulate and partially solve a toy model which
exhibits the simplest of the complexities of lattice formulations in
higher
than two dimensions, specifically which has several plaquettes glued to
the
same closed link (i.e. with $S_1$ topology) of the lattice.

We start with $U(N)$ and discuss the reduction to $SU(N)$ below.
An explicit basis $t^a_{i\bar j}$ for the fundamental representation of
the Lie
algebra, in terms of the basis matrices
$(e^{kl})_{i\bar j}= \delta^k_i \delta^l_{\bar j}$ is
\eqn\liedef{ T^{(ij)} \equiv e^{ij}+e^{ji};\qquad
 T^{[ij]} \equiv i(e^{ij}-e^{ji}); \qquad
 H^i \equiv e^{ii} - e^{i+1,i+1}; \qquad
 Q \equiv \sum_i e^{ii}.}
The Killing metric $g^{ab}=\trace t^a t^b$ is then
$g^{ab}=2\delta^{ab}$ for the off-diagonal generators,
$g^{ij}=2\delta^{ij}-\delta^{|i-j|,1}$
for the diagonal generators, and $g^{QQ}=N$.
The second Casimir $C_2 = g_{ab} t^a t^b$ then satisfies
$\trace C_2 = g_{ab} g^{ab} = \dim G$ so $C_2 = N$.

Group elements can be formed from the algebra as $U = \exp i\sum\tau^a
t_a$,
a general form which is particularly useful if we restrict $\tau^a$ to the
maximal torus, i.e. $a$ ranges over the diagonal generators.
Since we can diagonalize a unitary, conjugacy classes of the group are
just
points in the maximal torus.
We will also describe this torus as $U = \exp i\sum\theta^i e_{ii}$ or
$U_{i\bar j} = \delta_{i\bar j} z_i$.
$\int dU \sqrt{g }$ or simply $\int dU$ will be the left and
right-invariant
metric, normalized to $\int dU~= 1$.

Perhaps the simplest way to motivate a relation between $U(N)$
representation
theory and a fermionic theory is simply to quote the Weyl character
formula.
This gives the character
\eqn\char{\chi_R(\vec z) \equiv \Tr \prod_i z_i ^ {H_i} =
\tr \vec z^{~\vec H}}
of a representation with highest weight $\vec w$ as a ratio of finite sums
over
the Weyl group.
For $U(N)$, we can label a representation $R$ by a set of indices $n_i$,
$1\le
i \le N$ with $n_1>n_2>\ldots>n_N$, where $n_i-N+i$ is the number of boxes
in
the $i$'th row of the Young tableau, and $Q = \sum_i n_i - N(N+1)/2$ is
the
$U(1)$ charge.
Then
\eqn\weyl{\chi_{\vec n}(\vec z) =
{\det_{i,j} z_i^{n_j} \over \det_{i,1\le j\le N} z_i^j}.}

We notice that multiplying by the common denominator in the formula turns
a
character into an antisymmetric function of the $N$ variables, which is
normalized in the inner product $\int \prod_i d\theta_i$.
Thus this function $\psi_R(\vec z)$ can be regarded as the wave function
of a
quantum mechanical system of $N$ fermions.

Alternatively one could start from quantum mechanics on the $U(N)$ group
manifold.  Noting that characters span the class functions (functions
invariant
under $U\rightarrow gUg^{-1}$), one is just interested in the singlet
sector of
this quantum mechanics, which can be reduced to a system of $N$ free
fermions
following the analysis of hermitian matrix quantum mechanics by Brezin et.
al.
\ref\bipz{E. Brezin, C. Itzykson, G. Parisi and J.-B. Zuber,
Comm.~Math.~Phys.
59 (1978) 35.}.
Thus we start with a Hamiltonian
\eqn\qmham{H = \sum_a E^a E^a}
where $E^a = \Tr~ t^a U {\partial/\partial U}$ satisfy the conventions
above
and generate left rotations on $U$.
We restrict to the singlet sector by changing variables to
$U_{ij}=R_{ik} z_k R^{-1}_{kj}$
and taking wave functions independent of the unitary $R$.
Let $z_i=\exp i\theta_i$; wave functions are periodic under
$\theta_i\rightarrow\theta_i+2\pi$, and symmetric under permutation.
The invariant measure in these variables is
\eqn\haar{\sqrt{g} = |\Delta(z)|^2 = \tilde\Delta(z)^2}
where $\Delta(z)=\prod_{i<j}(z_i-z_j)$ and
$\tilde\Delta(z)=\prod_{i<j}\sin{\theta_i-\theta_j\over 2}
=\Delta(z)/\prod_i z_i^{(N-1)/2}$.
Thus
\eqn\qmhamtwo{H = -\sum_i {1\over\tilde\Delta^2}{d\over d\theta_i}
\tilde\Delta^2{d\over d\theta_i}.}
Following BIPZ we rewrite this as
\eqn\qmhamtwob{H = -\sum_i \left[ {1\over\tilde\Delta}{d^2\over
d\theta_i^2}
\tilde\Delta - {1\over\tilde\Delta}
\left({d^2 \tilde\Delta\over d\theta_i^2}\right)
\right] .}

For hermitian matrix quantum mechanics $\Delta$ was a Vandermonde and the
second term, thanks to a non-trivial identity, gave zero.  Here, after
similar
manipulations, the second term is found to equal $-N(N^2-1)/12$.
Thus, after redefining the wave functions by
$\psi\rightarrow\tilde\Delta\psi$,
we have a theory of $N$ free fermions on the circle.  The boundary
conditions
are also determined by this redefinition; they become periodic
(antiperiodic,
respectively) for $N$ odd (even).
An orthonormal basis for wave functions is Slater determinants
\eqn\slate{\psi_{\vec n} = \det_{i,j} z_i^{n_j}}
with energy $E = \sum_i n_i^2 - N(N^2-1)/12$.
The ground state has fermions distributed symmetrically about $n=0$, and
energy
zero, so the Fermi level $n_F = (N-1)/2$.

Since the Hamiltonian is just the second Casimir of the Lie algebra,
we might expect energy eigenstates to be associated with irreducible
representations.
Comparison with the Weyl character formula bears this out;
a direct proof follows from \stone, equation (4.8).

It is convenient to introduce a second quantized formalism
with operators $B^{+}_{-n}$ and $B_n$ creating and destroying the fermion
mode
$z^n$, and $\psi(\theta)=\sum_n e^{in\theta} B_n$.  Then
$H = \int d\theta \partial\psi^{+} \partial\psi$.

Another basis for class functions on the group manifold is products of
\eqn\wbasis{W_n = \Tr U^n = \sum_i z_i^n.}
In our second quantized formalism these are fermion bilinears.

It is now appropriate to take the large $N$ limit.
Many things simplify if we never consider operators $W_n$ with $n\sim N$,
because then fermions near the positive and negative Fermi surfaces
completely
decouple.
We can then speak of a relativistic fermi system, with complex chiral
left- and
right-moving fermions.  We should also speak of $U$ raising the
left-moving
(upper) fermions while lowering the right-movers, and $U^{-1}$ doing the
opposite (so these are exactly the two chirality sectors of Gross).
So, let
$b^{+}_{n} = B^{+}_{-n_F-\epsilon+n}$,
$b_{n} = B_{n_F+\epsilon+n}$,
$\bar b^{+}_{n} = B^{+}_{n_F+\epsilon-n}$,
$\bar b_{n} = B_{-n_F-\epsilon-n}$,
where $\epsilon=\half$ was just introduced to give Neveu-Schwarz
($n\in\IZ+\half$) modeing for all $N$.
The local operators $\psi(z)=\sum_{n\in\IZ+\half} z^{-n} b_n$,
$\psi^{+}(z)$,
$\bar\psi(\bar z)=\sum_{n\in\IZ+\half} \bar z^{-n} \bar b_n$, and
$\bar\psi^{+}(\bar z)$
now satisfy standard 2d field theory commutation relations.
The standard Fock vacuum ($b_n|0>= b^{+}_n|0>=0$ for $n>0$) corresponds to
the
identity representation, and higher representations can be built by acting
with
the bilinears $b^{+}_n b_m$ and
$\bar b^{+}_n \bar b_m$.
Of these, clearly the simplest are the $W_n$'s which become
\eqn\wrel{W_n = \Tr U^n =
\cint dz~ z^{-1-n} \psi^{+}(z) \psi(z) +
\cint d\bar z~ \bar z^{-1+n} \bar\psi^{+}(z) \bar\psi(z).}
We immediately recognize the operators here as the left- and right-moving
conformal field theory $U(1)$ currents, and the construction of the
$W_n$'s as
bosonization:
\eqn\walph{W_n \equiv \alpha_{-n} + \bar\alpha_{n}
= \int d\theta e^{in\theta}
\partial_\tau\phi(z=e^{n(\tau+i\theta)},\bar z=e^{n(\tau-i\theta)})}
defining the standard free boson oscillator expansion with
$[\alpha_m,\alpha_n]=\delta_{m+n,0}$ (resp. $\bar\alpha$), and
$[\alpha,\bar\alpha]=0$.
The charges $\alpha_0$ and $\bar\alpha_0$ count fermion numbers, which are
constant in our application, and we will simply define these to be zero.
The $W_n$ commute as operators, as they should.

The appearance of a two-dimensional auxiliary space in the problem
(sometimes
referred to as the ``world-sheet'' in the following, but only out of
habit) has
more to do with the number of degrees of freedom we are representing than
any
intrinsic two-dimensional nature of group theory.  As we will see below,
some
natural group-theoretic operations are non-local in this space.

A number of non-trivial results already follow.
First, we have a quantum field theory description of the well-known
Frobenius
relation between characters and symmetric polynomials.
A character $\chi_{\vec n}$ corresponds to a Fock basis state in a simple
way;
if one takes a ``chiral'' (in the sense of Gross) representation $R$,
i.e. in the tensor product of finitely many fundamental representations,
with Young tableau with $n_i$ boxes in the $i$'th row, $1\le i\le r$,
this corresponds to the state
\eqn\state{|\vec n> = \chi_{\vec n}(U)|0> =
b^{+}_{\epsilon-n_1} b^{+}_{\epsilon-n_2} \ldots b^{+}_{\epsilon-n_r}
b_{-\epsilon-1} b_{-\epsilon-2} \ldots b_{-\epsilon-r} |0>.}
An alternate basis for class functions is
\eqn\class{\prod_i (\Tr U^i)^{\sigma_i}}
(in terms of the $z_i$ these functions give a basis for the symmetric
polynomials)
which will correspond to the state
\eqn\clstate{|\sigma> = \prod_i W_i^{\sigma_i} |0>.}
Orthonormality of the characters gives us
\eqn\res{\chi_{\vec n}(U)|0> = |\vec n> = \sum_\sigma <\vec n|\sigma>
W_i^{\sigma_i} |0>.}

This is also true for products of ``chiral'' and ``anti-chiral''
representations, and in explaining this it is also appropriate to discuss
the
reduction to $SU(N)$ from $U(N)$.
On a group element this is accomplished by constraining
$z \equiv \prod_i z^i = 1$.
Equivalently, we can restrict the class functions we consider to those
with
zero $U(1)$ charge.  Since $Q=\sum_i z_i d/dz_i$, we can multiply a
function
with charge $q$ by $z^{-q/N}$ to cancel its charge.
In the fermi language we would have $N$ twist sectors defined by their
charge
under the action of the $Z_N$ center of the group.
In the large $N$ limit this structure becomes irrelevant, and we can take
our
factor $\epsilon$ to depend on the sector in a way which gives all sectors
Neveu-Schwarz boundary conditions.
In principle we need to keep track of this in rewriting our Hamiltonian in
terms of the relativistic states, which we will do below.
However, a simpler procedure will be to explicitly subtract the
contribution
due to the $U(1)$ charge, which is still accessible after the rewriting,
using
$Q=L_0-\bar L_0$.

Otherwise, this modification has no effect, and we can check for example
that
$\Tr U \Tr U^{-1} |0> = (\alpha_{-1}\bar\alpha_{-1}+1)|0>$ contains the
singlet
and adjoint irreducible representations.
The highest component in the product of ``chiral'' and ``anti-chiral''
representations is then clearly the one in which the left- and
right-moving
fermions are excited independently, and the formula \res~ is again true,
where
we take bosonic states
$\sum_{\sigma,\bar \sigma} c_{\sigma,\bar \sigma} W_i^{\sigma_i}
W_{-i}^{\bar\sigma_i}|0>$
with $\sum_i i \sigma_i = L_0$ (the number of chiral boxes) and
$\sum_i i \bar\sigma_i = \bar L_0$.

Another application of this is the integration of class functions, which
is
just expectation values of products of operators.
For example,
\eqn\intex{\eqalign{\int dU (Tr U^{-2})(Tr U)^2  &=
<0|(\alpha_2+\bar\alpha_{-2})(\alpha_{-1}+\bar\alpha_{1})^2|0>\hfill\cr
&= 0}}
to all orders in $1/N$.
(and for finite $N>2$, by going back to the non-relativistic fermions.)

As we saw earlier, the quantum mechanical Hamiltonian is the second
Casimir,
which is not the relativistic Hamiltonian $L_0+\bar L_0$.
In terms of relativistic fermions it is
\eqn\rham{\eqalign{H_{U(N)} &= \sum_{n\in\IZ+\half} (n_F+\epsilon-n)^2
(b^{+}_{-n}b_n + \bar b^{+}_{n}\bar b_{-n}) - E_0\hfill\cr
&= NL_0 + N \bar L_0 + \sum_{n\in\IZ+\half} n^2
:b^{+}_{-n}b_n + \bar b^{+}_{-n}\bar b_{n}:\hfill\cr
&= NL_0 + N \bar L_0 + \cint dz~z^2 :\partial\psi^{+}\partial\psi:
+ \cint d\bar z~\bar z^2 :\bar \partial\bar \psi^{+}\bar \partial\bar
\psi:}}
(we know that the vacuum energy in this ground state is zero).
For $SU(N)$ we would subtract from this
\eqn\uone{H_{U(1)} = {1\over N}(L_0-\bar L_0)^2.}

The bosonization of this Hamiltonian is very well known in the context of
matrix models, as it is just the Das-Jevicki-Sakita Hamiltonian
governing the dynamics of the eigenvalue density in hermitian matrix
quantum
mechanics.  Of course we have just retraced the steps of Gross and
Klebanov
leading to free relativistic fermions in that problem.
Using abelian bosonization as in
\ref\gk{D.J. Gross and I. Klebanov, Nucl. Phys. B344 (1990) 475},
we reproduce
\eqn\boseham{H = N L_0 + N \bar L_0
+ {1\over 3}\cint dz~z^2 :(\partial\phi)^3:
- {1\over 3}\cint d\bar z~\bar z^2 :(\bar\partial\phi)^3:~.}

The cubic interaction term in this Hamiltonian is quite natural, as we
could
see by considering the action of our original \qmham~ on states \clstate~
--
it would contain terms preserving the ``string number'' (number of
traces), as
well as terms joining or splitting strings in higher order in $1/N$.

A direct application of quantum mechanics on a group manifold is to
two-dimensional Yang-Mills theory on the cylinder, where the wave function
is a
function of the gauge holonomy around the non-contractible loop, and
quantum
mechanical time corresponds to the area of the cylinder.
The large $N$ limit is taken with gauge coupling $g^2 = \tau/N$, $\tau$
fixed,
so time evolution is generated by an $O(N^0)$ Hamiltonian with
an $O(1/N)$ interaction term.
The torus partition function is easy in this formalism as it is just the
trace
of the heat kernel, so it is the torus partition function in our conformal
field theory.  The leading $N^0$ term is reproduced by free field theory,
while
the interaction terms in the Hamiltonian will generate $1/N$ corrections.
It is amusing that the original Yang-Mills states could be thought of as
strings winding around the cylinder with $\Tr U^n$ creating an $n$-winding
string, and that we have reformulated this as an operator creating a state
on
another cylinder of momentum $n$.

To extend this to arbitrary genus closed surfaces we need in addition the
wave
function for the disk and the three-holed sphere.
We might expect to be able to represent them as simple conformal field
theory
states, as is done in string field theory.
In particular the zero-area limits of these vertices satisfy simple
constraints
in the bosonic language.
First, the zero-area disk has trivial holonomy, or the wave function
$\psi=\delta(U)$.
In other words $\Tr U^n = N ~~\forall n\ne 0$, or
\def\zeroname {D_0}
\eqn\diskcon{(\alpha_n + \bar\alpha_{-n} - N)|\zeroname> = 0.}
These constraints are easily solved:
\eqn\diskzero{|\zeroname> =
\exp ~-\sum_{n\ge 1}{1\over n}(\alpha_{-n}\bar\alpha_{-n} - N \alpha_{-n}
- N \bar\alpha_{-n}) ~~|0>.}
This state can be used to calculate the dimension of a representation:
\eqn\dimr{\dim R_{\vec n} = <\zeroname|\vec n>.}
(try for example the characters $\half((\Tr U)^2+\Tr U^2)$.)

Another characterization of this state is through boundary conditions of
the
fields -- we put a boundary $\tau=0$ with
$\partial\phi/\partial\tau = N\delta(\theta)$.
This is equivalent to taking Neumann boundary conditions, and inserting
the
operator $:\exp iN\phi(0):$.
In the fermi language this is a rather singular state, which is defined by
the
boundary condition that all the fermions (eigenvalues) are at $z=1$.
In the non-relativistic formalism, we can produce it by taking the o.p.e.
of
$N$ fermions, resulting in the state
$\prod_{i=0}^{N-1} \partial^i\psi~~|0>.$
Taking the inner product of (for example) a character with this state will
reproduce the calculation of the dimension of a representation by taking
the
limit of all $z_i\rightarrow 1$ in the Weyl character formula using
l'H\^opital's rule.

Neither form is very easy to work with, and an attempt to calculate the
leading
$O(N^2)$ term in the QCD${}_2$ sphere free energy will illustrate this.
This is simply
\eqn\free{ \exp N^2 F_0 = <D_0|e^{-\tau H}|D_0>}
and in our bosonic formulation this looks like a calculation of the
partition
function on a cylinder with Neumann boundary conditions, with an
interaction
suppressed by $1/N$, and two local sources $e^{N\phi(0)}$ on the
boundaries.
What makes this complicated is that the source has an $N$ in the exponent,
which compensates the $1/N$ suppression of the interaction.
After going to the Lagrangian and rescaling $\phi\rightarrow N\phi$, the
action
has an $N^2$ in front of every term, showing both that the leading term in
the
free energy is $O(N^2)$ and that we can get it by solving a classical
problem,
however the classical problem is non-linear.
This is of course the same problem that Gross
\ref\gr{D. Gross, Princeton preprint PUPT-1356, hep-th/9212149}
found in directly performing the sum; one must keep the $O(1/N)$
subleading
term in the Casimir since it can combine with $N$'s from the dimension
formula.
The integrable nature of the bosonic theory and relation to free fermions
should make this problem solvable;
however we leave the full treatment of this for future work.

The other natural object for QCD${}_2$ is the ``three-space vertex'', i.e.
we
can build a Riemann surface of higher genus by connecting cylinders with a
vertex which turns one $S_1$ into two, identifying the product of the
holonomies of the resulting $S_1$'s with that of the original one.
In terms of wave functions this is simply
\eqn\vert{\eqalign{
 \psi_2(V) \otimes \psi_3(W) &= \int dU \psi_1(UVU^{-1}W) \hfil\cr
&= \sum_R {1\over \dim R} \chi_R(V) \chi_R(W) <R|\psi_1>, }}
which can certainly be written formally as a sum over fermionic states.
Some idea of how unnatural this is in our conformal field theory formalism
can
be gained by considering that one property of this vertex is that
shrinking one
of the $S_1$'s to zero, i.e. gluing the state $|D_0>$, produces the
identity
operator on the product of the remaining $S_1$'s.
Since $|D_0>$ was a finite-size Neumann boundary, the vertex is highly
non-local in our auxiliary space.
Another disturbing picture of this sort follows from considering the
vertex
with two $S_1$'s identified, producing an operator which adds a zero-area
handle, $|H_0>$.  This is a fairly benign object in QCD${}_2$ itself due
to the
area-preserving diffeomorphism invariance, but here must for example
reproduce
the torus partition function, also naturally thought of as a path integral
on
the torus in the present formalism, as
$<H_0|e^{-\tau H}|D_0> = Z_1(\tau)$.

A conclusion one might draw from these pictures is that the present
formalism
is very unnatural for QCD${}_2$ problems.
Although there is certainly truth to this, there are other interesting
operations which are quite simple in this formalism and which have
applications
in other toy models of large $N$ gauge theory.
A good example is multiplication of wave functions, which we could use for
example to compute
\eqn\nplaq{Z_n = \int dU Z^n_{hk}(U).}
Expressing the heat kernel action in terms of characters, we would need
\eqn\mchar{\int dU \chi_R(U) \chi_S(U) \chi_T(U) = N_{RST},}
where $N_{RST}$ is the number of singlets in the tensor product of these
irreducible representations.
A number of formulas exist for these coefficients
\ref\gr{W. Fulton and J. Harris, Representation theory : a first course,
Springer-Verlag, 1991}
but are not convenient for performing sums of the sort we want.
However in the present formalism, the operation of multiplying wave
functions
is natural, and local on the world-sheet.
The operation of multiplying $n$ wave functions is representable by a
``$n$-string vertex'' which is characterized by the following property:
\eqn\wchar{<\psi_1|<\psi_2|\ldots \Tr U^k_i |V_n> =
\int dU~ \Tr U^k \prod_i \psi_{i}(U)}
is the same no matter which wave function we multiply by $\Tr U^k$.
In conformal field theory terms this means that we have a boundary on
which $n$
cylinders meet, and the boundary condition
\eqn\wver{\partial_\tau\phi_i(z)=\partial_\tau\phi_j(z) \qquad\forall
i,j.}
This condition only couples mode $n$ to $-n$ and can be translated
\ref\mrd{M. Douglas, to appear.}
into an oscillator expression for the vertex
\eqn\vert{|V_n> = \exp \sum_{i\ne j}
\sum_{n\ge 1} {1\over n} \alpha_{-n}^{(i)} \bar\alpha_{-n}^{(j)} ~~|0>.}
We have verified that contracting $|V_3>$ with a general state
$|1>=\prod_i
(\Tr U^i)^{\sigma_i}|0>$ produces an operator which is exactly
multiplication
by $\prod_i (\Tr U^i)^{\sigma_i}$.

Using this vertex, a number of models with multiple plaquettes joined on a
link
can be solved.  Of course the link must be closed, so that we only need
the
partition function on the incident plaquettes as a function of each
holonomy
separately, and as we mentioned our results are not very explicit for
plaquettes forming higher-genus two-dimensional surfaces.
An example of a model we can solve is to take $D$ cylinders, identify one
end
of each cylinder as above, and identify the other ends separately.
This is the partition function
\eqn\eko{Z_D = <V_D| \exp -\sum_i \tau_i H_i |V_D>.}

\nref\ekm{T. Eguchi and H. Kawai, Phys. Rev. Lett. 48 (1982) 1063.}
In fact this is (a slight generalization of) a limit of the Eguchi-Kawai
reduced model of QCD \ekm in $D$ dimensions, where we keep only the
electric
field plaquettes
$F_{0i} = U_0 U_i U_0^{-1} U_i^{-1}.$
To see this, first imagine compactifying time; the plaquette $F_{0i}$ then
becomes a cylinder.  Gluing these plaquettes to the one time-like link
then
produces the above partition function with one $\tau$ set to zero.
Of course throwing away the magnetic field removes any physical connection
with
higher-dimensional QCD; nevertheless this is an exercise which gives the
number
of degrees of freedom in the theory, and provides a toy model on which
generalizations of \gt~ can be tested.

This model has no $N^2$ term in its free energy.  Therefore the
complications
of the sphere partition function are absent, and the $N^0$ term is a free
oscillator problem.  Corrections in $1/N$ can be generated by insertions
of the
cubic interaction Hamiltonian.  We quote the final result at $O(N^0)$
(with
$q_i = \exp -\tau_i$):
\eqn\result{Z_{3} = \prod_{n\ge 1}
(1-\sum_{i<j}q_i^n q_j^n + 2 q_1^n q_2^n q_3^n)^{-1}
(1-\sum_{i<j}q_i^n q_j^n - 2 q_1^n q_2^n q_3^n)^{-1}.}
and
\eqn\resultn{Z_{D} = \prod_{n\ge 1}
(1-\sum_{a=2}^{D} (a-1)E_a(q^n))^{-1}
(1-\sum_{a=2}^{D} (a-1)E_a(-q^n))^{-1} }
where $E_a(q_i) = \sum_{i_1<\ldots<i_a} q_{i_1} q_{i_2} \ldots$ are the
elementary symmetric polynomials.

Setting $q_1=q_2=q$, $q_3=1$ in the $D=3$ answer gives the EK partition
function in this case,
\eqn\ekr{Z_{EK3} = \prod_{n\ge 1} (1-q^n)^{-2} (1+q^n)^{-1} (1-3q^n)^{-1}
{}~~+ O(1/N^2).}

The $(1-3q^n)$ term demonstrates the expected exponential growth in the
density
 of states for a string theory.  In particular one should not think of the
states in our CFT description as having a one-to-one correspondance with
states
in the EK theory.  Making direct progress on the model with the magnetic
field
term almost certainly requires a description of these states.  Of course
one
can write a simple basis like $\Tr U^n$, $\Tr U^n V^m$, $\Tr U^n V^m U^p
V^q$,
and so forth, (and elementary combinatorics using this basis reproduces
the
above $O(N^0)$ result), but one should not expect a field theory
representation
of this basis.  This is why we are looking for a QCD string.
\vskip 0.5in
This project was initiated in conversations with J. Polchinski and A.
Strominger at the ITP Non-perturbative String Physics workshop.
I thank J. Polchinski for pointing out \diskcon.
I would also like to acknowledge enlightening discussions with T. Banks,
J.
Cohn, and P. Ginsparg.

The reader is also directed to a simultaneous and independent
work by J. Minahan and
A. Polychronakos, hep-th/9303153, with substantial overlap with the
present
work.

This work was supported in part by DOE grant DE-FG05-90ER40559, NSF grants
PHY-9157016 and PHY89-04035, and the Sloan Foundation.

\listrefs
\end